\documentclass[%
 reprint,
 amsmath,amssymb,
 aps,
]{revtex4-2}

\usepackage{hyperref}
\usepackage{graphicx}
\usepackage{lineno}
\usepackage{dcolumn}
\usepackage{bm}
\usepackage{epstopdf}
\usepackage{amsmath,amssymb}

\newcommand{\vone}{$v_{1}$}
\newcommand{\veta}{$v_{1}(\eta)$}
\newcommand{\snn}{$\sqrt{s_{NN}}=$}
\newcommand{\psione}{$\Psi_{1}$}
\newcommand{\dndeta}{$\frac{\mathrm{d}N}{\mathrm{d}\eta}$}
\newcommand{\vpt}{$v_{1}(p_{T})$}
\newcommand{\pt}{$p_{T}$}
\newcommand{\vz}{$V_{z}$}

\newcommand{\deltaphi}{$\phi-\Psi_{1}^{\mathrm{TPC}}$}
\newcommand{\psitpc}{$\Psi_{1}^{\mathrm{TPC}}$}

\DeclareUnicodeCharacter{2212}{-}
\begin{document}

\preprint{APS/123-QED}

\title{Measurement of directed flow in Au+Au collisions at \snn\ 19.6 and 27 GeV with the STAR Event Plane Detector}

\author{STAR Collaboration}
\date{\today}

\begin{abstract}
In heavy-ion collision experiments, the global collectivity of final-state particles can be quantified by anisotropic flow coefficients ($v_n$). The first-order flow coefficient, also referred to as the directed flow (\vone), describes the collective sideward motion of produced particles and nuclear fragments in heavy-ion collisions. It carries information on the very early stage of the collision, especially at large pseudorapidity ($\eta$), where it is believed to be generated during the nuclear passage time. Directed flow therefore probes the onset of bulk collective dynamics during thermalization, providing valuable experimental guidance to models of the pre-equilibrium stage. In 2018, the Event Plane Detector (EPD) was installed in STAR and used for the Beam Energy Scan phase-II (BES-II) data taking. The combination of EPD ($2.1 <|\eta|< 5.1$) and high-statistics BES-II data enables us to extend the \vone\ measurement to the forward and backward $\eta$ regions. In this paper, we present the measurement of $v_{1}$ over a wide $\eta$ range in Au+Au collisions at $\sqrt{s_{NN}}=$ 19.6 and 27 GeV using the STAR EPD. The results of the analysis at \snn\ 19.6 GeV exhibit excellent consistency with the previous PHOBOS measurement, while elevating the precision of the overall measurement. The increased precision of the measurement also revealed finer structures in heavy-ion collisions, including a potential observation of the first-order event-plane decorrelation. 
Multiple physics models were compared to the experimental results. Only a transport model and a three-fluid hybrid model can reproduce a sizable \vone\ at large $\eta$ as was observed experimentally. The model comparison also indicates \vone\ at large $\eta$ might be sensitive to the QGP phase transition. 

\end{abstract}

\maketitle


\section{\label{sec:level1}Introduction}
Nuclear collisions at ultra-relativistic energies create a deconfined system of partons in the midrapidity region~\cite{Heinz:2000bk,STAR:2005gfr,PHENIX:2004vcz,PHOBOS:2004zne,BRAHMS:2004adc}. The so-called quark-gluon plasma (QGP) represents the state of QCD matter under conditions present in the first few microseconds after the Big Bang. Studying the properties of the QGP and the transition from confined to deconfined color matter is the primary purpose of the experimental program at the Relativistic Heavy Ion Collider (RHIC) at Brookhaven National Laboratory.

The QGP created at RHIC can be understood remarkably well when modeled as a hydrodynamic system~\cite{Gale:2013da}, providing access to the equation of state (EoS) and transport coefficients~\cite{Heinz:2009xj,Heinz:2013th} of the QCD matter under extreme conditions. Measurements of anisotropic flow~\cite{Voloshin:2008dg} have proven particularly useful to extract both the QGP properties as well as the nontrivial initial conditions~\cite{Alver:2010dn,Alver:2010gr,Singha:2016mna}. The hot zone formed in any collision is geometrically anisotropic, leading to anisotropic pressure gradients and hydrodynamic response. Measured flow coefficients, $v_n$, quantify the azimuthal anisotropy of particle emission relative to an event-plane $\Psi$:
\begin{equation}
    \frac{\mathrm{d}N}{\mathrm{d}\phi}=\frac{N}{2\pi}\{1+\sum_{n=1}^{\infty}2v_{n}\cos{[n(\phi-\Psi_{n})]}\}. 
\end{equation}
The event-plane angle is measured as:
\begin{equation}
    \Psi_{n}=\frac{1}{n}\arctan{\frac{\sum_{i}w_{i}\sin{(n\phi_{i})}}{\sum_{i}w_{i}\cos{(n\phi_{i})}}},
\end{equation}
where $w_i$ is the weight assigned to the $i^{\text{th}}$ particle, and the sums run over all the particles that are used to calculate the event plane. Since the event plane is measured with a finite number of particles, the anisotropic flow measured relative to the event plane needs to be corrected by the event-plane resolution~\cite{Poskanzer:1998yz}. 



The thermodynamic variables quantifying the state of matter are temperature ($T$) and baryon chemical potential ($\mu_B$)~\cite{Sorensen:2023zkk}. 
At $\mu_{B}=0$, first-principle lattice QCD calculations have established the transition between the QGP and hadron gas to be a crossover transition at the critical temperature $T_{c}=154\pm9$ MeV~\cite{HotQCD:2014kol}. At finite $\mu_{B}$, QCD-based models predict a first-order phase transition and the existence of a critical point at the end of the first-order phase transition line~\cite{Ejiri:2008xt,PhysRevC.79.015202}. Since the first-principle lattice calculations of QCD EoS are only stable at very low values of $\mu_B$~\cite{Guenther:2020jwe}, insights on the transition from color-confined to -deconfined states will depend crucially on experimental measurements and detailed modeling. 

In 2010, RHIC embarked upon the Beam Energy Scan (BES) program~\cite{STAR:2010vob}, with a goal to 
further explore the QCD phase diagram at finite $\mu_{B}$. The first stage of the RHIC BES program was carried out during 2010 and 2014. Au+Au data were collected by the STAR experiment at six energies (\snn\ 62.4, 39, 27, 19.6, 14.5, 11.5 and 7.7 GeV). As the collision energy is reduced, the QGP is doped with more quarks than antiquarks, thus reaching a higher $\mu_B$~\cite{Busza:2018rrf,Aprahamian:2015qub,Cleymans:2006zz}. Several measurements from BES-I suggest that a phase transition occurs in collisions near $\sqrt{s_{NN}}\approx 20$~GeV, an order of magnitude lower than the top RHIC energy~\cite{STAR:2014clz,STAR:2020tga}. 
From 2018 to 2021, RHIC conducted the second phase of the BES, focusing on the energies around the expected location of the transition energy: \snn\ 27, 19.6, 17.3, 14.6, 11.5, 9.2 and 7.7 GeV.

At BES energies, the colliding system is less boosted compared to the top RHIC energy, so a significant amount of evolution happens before nuclei have completely passed through each other. Therefore any dynamical model must treat the full three-dimensional system in detail and the simulation of the pre-equilibrium stage becomes important.  Furthermore, the baryon doping of the region of interest, requires transporting baryon number from $y=y_{\rm beam}$ to the midrapidity~\cite{Videbaek:2009zy,BRAHMS:2003wwg}; the mechanism of this so-called ``baryon stopping'' remains a topic of intense research~\cite{Brandenburg:2022hrp,Lewis:2023nmd,Lv:2023fxk,Dong:2023zbu}. A satisfactory understanding of heavy-ion collisions in the BES transition region will thus require modeling the dynamics over the entire rapidity range.

The first-order flow coefficient (\vone), also referred to as the directed flow, quantifies the sideward motion of produced particles and nuclear fragments as a function of (pseudo)rapidity. 
It vanishes by symmetry in a purely boost-invariant model, in which the system is independent of the space-time rapidity. It is particularly interesting at BES energies, as it connects the longitudinal and transverse dynamics, manifestly probing the three-dimensional nature of the system's evolution. Directed flow is also sensitive to the early stage of the collision, especially in the fragmentation region, where it is believed to be generated during the nuclear passage time 
($\sim~1~\mathrm{fm}/c$)~\cite{Sorge:1996pc,Herrmann:1999wu}. 
Directed flow therefore probes the onset of bulk collective dynamics during thermalization, providing valuable experimental guidance to models of the pre-equilibrium stage~\cite{STAR:2010vob}. Model studies have indicated that directed flow is sensitive to the shear viscosity of the hot QCD matter~\cite{Becattini:2015ska}. Furthermore, directed flow has demonstrated strong constraining power on the initial baryon stopping and can serve as a probe for the EoS in heavy-ion collisions~\cite{Ivanov:2017xbd,Du:2022yok,Jiang:2021foj}. Currently, most model studies at BES energies focus around midrapidity due to both the lack of experimental data in the forward (backward) region and an insufficient understanding of particle production 
in the fragmentation region. However, past measurements have indicated that the directed flow signal is most pronounced at the forward(backward) (pseudo)rapidity~\cite{STAR:2005btp,STAR:2003xyj,STAR:2004jwm,STAR:2005btp,STAR:2008jgm}. Therefore, any sensitivity of \vone\ to the initial state, transport coefficients, or the EoS may be more evident at large (pseudo)rapidities. 
The measurement of directed flow over a wide pseudorapidity range will thus offer valuable constraints on the three-dimensional initial state and evolution of the heavy-ion collision.

In this paper, we use the STAR Event Plane Detector~\cite{Adams:2019fpo} (EPD) to measure directed flow at $\sqrt{s_{NN}}=19.6$ and 27 GeV, from $2.1<|\eta|<5.1$. The EPD is part of the STAR BES-II upgrade. It is a hit detector and was originally designed to measure event-plane angles ~\cite{Adams:2019fpo}.
It is challenging to use the EPD as the particles-of-interest (PoI) region in the flow measurement because it cannot count the exact number of particles traversing each tile per event. Therefore, we developed an entirely new method to ensure the accuracy of this measurement. This method will help STAR to extend flow measurements to a wide pseudorapidity range at all the BES-II energies.


This paper is organized as follows. The STAR detector and the BES-II data set are introduced in Sec.~\ref{sec:level2}. The details of how the reference is chosen and how the event plane is calculated are discussed in Sec.~\ref{sec:level3_1}. Our new approach to measure $v_{1}$ in the EPD acceptance is discussed in Sec.~\ref{sec:level3_2}. The correction for the STAR material effect is discussed in Sec.~\ref{sec:level3_3}. Systematic checks are discussed in Sec.~\ref{sec:level4}. In Sec.~\ref{sec:level5}, we present \veta\ measured from $2.1<|\eta|<5.1$ in Au+Au collisions at \snn\ 19.6 and 27 GeV. Comparisons with various models are also discussed. Finally, the analysis results are summarized in Sec.~\ref{sec:level6}.

\section{\label{sec:level2}Experiment}

\subsection{\label{sec:level2_1}STAR detector subsystems}
\begin{figure}[htb]
    \centering
    \includegraphics[width=1\columnwidth]{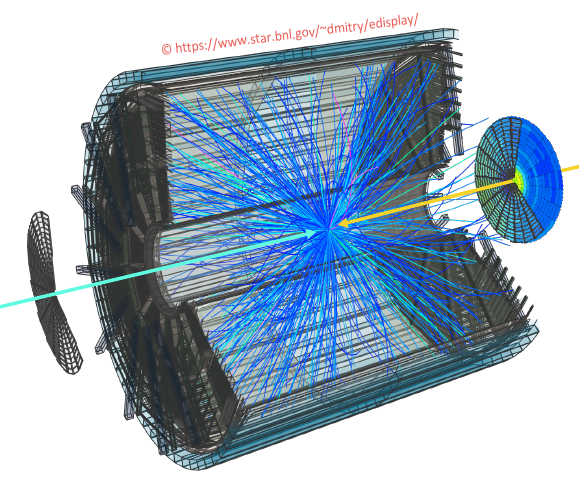}
    \caption{A sketch of an event recorded by the TPC and the EPD}
    \label{fig:star_illustration}
\end{figure}

\begin{figure}
    \centering
    \includegraphics[width=0.8\columnwidth]{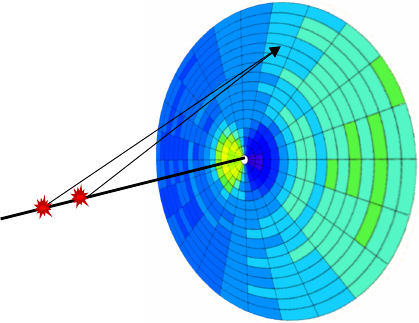}
    \caption{An illustration of the EPD tiles lighting up in a single event. In this case, the right side of the EPD wheel is mainly hit by produced particles from the fireball, while the left side of the EPD wheel is mainly hit by the nuclear fragments. The pseudorapidity ($\eta$) range of an EPD tile depends on the primary vertex position. The EPD acceptance is $2.1<|\eta|<5.1$ when the collision happens at the center of the TPC.}
    \label{fig:epd_illustration}
\end{figure}
The main subsystems of STAR used in this paper are the Time Projection Chamber (TPC) and the EPD. Figure~\ref{fig:star_illustration} shows a collision event recorded by these two detectors. The TPC is 4.2 m long and 4 m in diameter. It is used to detect charged particles within the pseudorapidity range of $|\eta|<1$, with a full 2$\pi$ azimuthal ($\phi$) coverage~\cite{Anderson:2003ur}. In this analysis, we used TPC tracks with $0.15<p_{T}<2.0$ GeV/$c$. The TPC is also used to reconstruct the primary vertex position of each event along the beam direction ($V_{z}$) and its radial distance from the $z$ axis ($V_{r}$). In this analysis, we require events with $|V_{z}|<40$ cm and $V_{r}<1$ cm. 
In order to reduce contamination from the interactions between the primary particles and the detector material, as well as the secondary particles from weak decays, we require tracks reconstructed in the TPC to have a distance of closest approach (DCA) to the primary vertex of less than 3 cm. We also require each track to have at least 16 ionization points in the TPC. 

The EPD consists of two scintillator wheels with high granularity located at $\pm3.75$ m from the center of the TPC along the beam direction. Each EPD wheel is composed of twelve “supersectors” that subtend 30 degrees in azimuth and each supersector is divided into 31 tiles~\cite{Adams:2019fpo}. As shown in Figure~\ref{fig:epd_illustration}, the peusorapidity range of an EPD tile depends on the primary vertex position, the $\eta$ and $\phi$ of a tile are determined by a straight line between the primary vertex and a random point on the tile. When the primary vertex is at the center of the TPC, the EPD acceptance is $2.1<|\eta|<5.1$. 
When a minimum-ionizing particle (MIP) traverses an EPD tile, the tile absorbs part of its energy and emits photons. The wavelength-shifting fibers wired in the EPD tile will transport the light to a silicon photomultiplier (SiPM). Signals from the SiPM are then amplified and sent to the STAR digitizing and acquisition system and are eventually recorded as ADC values. The non-MIP will deposit more energy than the MIP in the EPD tile, but the amount of non-MIPs is negligible compared to the amount of MIPs at BES energies.  As a pre-shower scintillator detector, the EPD cannot reconstruct charged tracks like the TPC does. However, the number of MIPs traversing each EPD tile averaged over all the events can be probabilistically determined. The extraction of averaged MIPs will be discussed in detail in Sec.~\ref{sec:level3_2}.

\subsection{\label{sec:level2_2} Data sets}
This analysis utilizes data from Au+Au collisions at \snn\ 27 GeV collected in 2018 and Au+Au collisions at \snn\ 19.6 GeV collected in 2019 as part of the BES-II program. 
STAR collected minimum-bias events by requiring the coincidence of signals on both sides of the interaction region from the Zero Degree Calorimeters (ZDCs), or the Vertex Position Detector (VPD), or the Beam Beam Counter (BBC). By excluding outliers in the correlation between the number of TPC tracks and the number of those tracks with a matched hit in the Time of Flight (TOF) detector, we are able to detect and remove out-of-time pile-up in roughly $0.02\%$ of these minimum-bias events. This is possible since the TOF is a fast detector and does not detect out-of-time pile-up events, unlike the TPC. 
%
We also require at least one TPC track matched to the TOF for selecting good events. After these event cuts together with the primary vertex cuts mentioned in Sec.~\ref{sec:level2_1}, 320 M events are available for the analysis at \snn\ 27 GeV and 260 M events are available at \snn\ 19.6 GeV, both for 0-80$\%$ centrality. The centrality classes are defined based on the charged-particle multiplicity ($N_{\mathrm{ch}}$) distribution in the TPC within the pseudorapidity window of $|\eta|<0.5$. Such distributions are fit to Monte Carlo (MC) Glauber simulations after correcting for the luminosity and acceptance variation as a function of $V_z$. The detailed procedure to obtain the simulated multiplicity using the MC Glauber is similar to that described in Ref.~\cite{STAR:2012och}. 

\section{\label{sec:level3}Method of Analysis}
\subsection{\label{sec:level3_1}Event-Plane Reconstruction}
Anisotropic flow reveals the global collectivity of the final state particles. However, the measurement of it can be contaminated by non-collective or nonflow correlations from various sources.
They include the momentum-conservation effect, quantum statistics, resonance decays, jet or minijet fragmentation, among others~\cite{Poskanzer:1998yz,Borghini:2002mv}. Therefore, the reference particles used to determine the event plane need to be carefully selected to suppress those nonflow effects. In this analysis, the TPC was chosen as the reference to suppress the momentum-conservation effect and the short-range correlations~\cite{Borghini:2002mv}. 
During the RHIC run in 2018, one of the 24 TPC sectors was used to commission the inner TPC (iTPC) sector and the data from this sector were not used for physics analyses. The loss of tracks due to the iTPC sector leads to a region of depletion in the $\eta-\phi$ acceptance map. Commonly, such issues are resolved by implementing a $\phi$-weighting. Nevertheless, when the collision vertex is displaced considerably from the center of the TPC, for instance, $V_z$ in the vicinity of $-$40 cm and $\eta$ close to $-$1, the regions of depletion are too prominent to be corrected by the $\phi$ weighting. Therefore, only tracks within $|\eta|<0.8$ were used at \snn\ 27 GeV. In run 19, there is no such issue, so, tracks within $|\eta|<1.0$ were used at \snn\ 19.6 GeV.

A weight of $w(\eta) = -\eta$ was applied to the TPC tracks since directed flow is odd in pseudorapidity. Without this weight, the asymmetry of the collision would result in a resolution of zero for the first-order event plane reconstructed from TPC ($\Psi_{1}^{\rm TPC}$). Another track weight based on the $\eta$ and $\phi$ of the TPC tracks was also used to achieve a uniform event-averaged $dN/d\phi$ distribution and a symmetric $dN/d\eta$ distribution around mid-rapidity.

\begin{figure}
    \centering
    \includegraphics[width=\columnwidth]{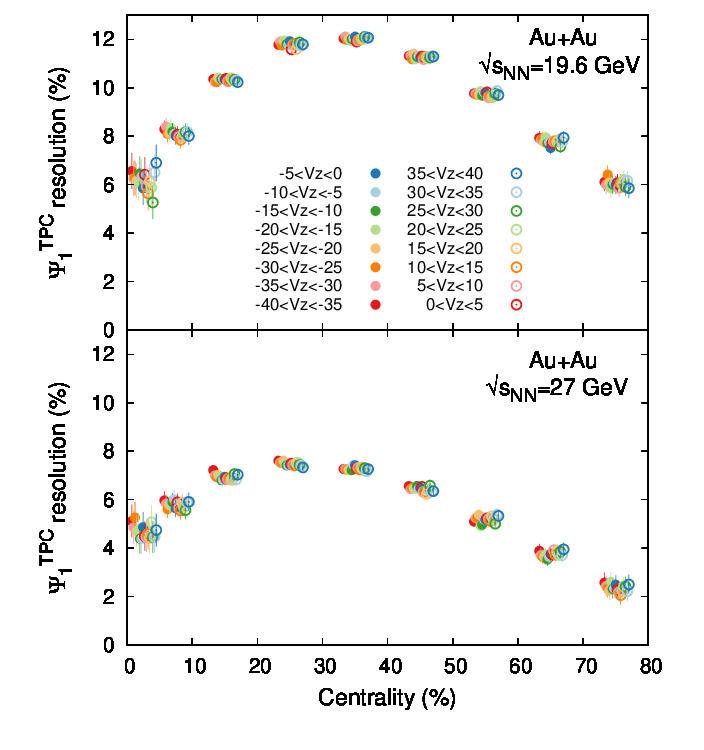}
    \caption{$\Psi_{1}^{\mathrm{TPC}}$ resolution as a function of centrality for 16 $V_{z}$ bins; the resolution is calculated by the three sub-event method. The resolution differences between two collision energies mainly come from the different pseudorapidity regions used. The data points are shifted along the x-axis for a clearer presentation.}
    \label{fig:tpc_psi1_res}
\end{figure}

The resolution of the TPC event plane is calculated by the ``three sub-event method'':
\begin{equation}\label{eq:tpc_resolution}
    R_{1}^{\mathrm{TPC}}=\sqrt{\frac{\langle\cos{(\Psi_{1}^{\mathrm{TPC}}-\Psi_{1}^{\mathrm{EPDE}})}\rangle\langle\cos{(\Psi_{1}^\mathrm{TPC}-\Psi_{1}^\mathrm{EPDW})}\rangle}{\langle\cos{(\Psi_{1}^\mathrm{EPDE}-\Psi_{1}^\mathrm{EPDW})}\rangle}},
\end{equation}
where $\Psi_{1}^{\mathrm{EPDE}}$ is the event-plane angle measured by the East EPD ($-5.1<\eta<-2.1$) and $\Psi_{1}^{\mathrm{EPDW}}$ is the event-plane angle measured by the West EPD ($2.1<\eta<5.1$). All the event-plane distributions have been flattened by the shifting method~\cite{Poskanzer:1998yz} to further remove the acceptance correlations from an imperfect detector. Figure~\ref{fig:tpc_psi1_res} shows the $\Psi_{1}^{\mathrm{TPC}}$ resolution as a function of centrality at \snn\ 19.6 and 27 GeV for sixteen \vz\ bins between $-$40 to 40 cm. The resolution peaks at mid-centrality at both energies, and no \vz\ dependence was observed. 

\subsection{\label{sec:level3_2}Extraction of $v_{1}$}
\begin{figure}[t]
    \centering
    \includegraphics[width=1\columnwidth]{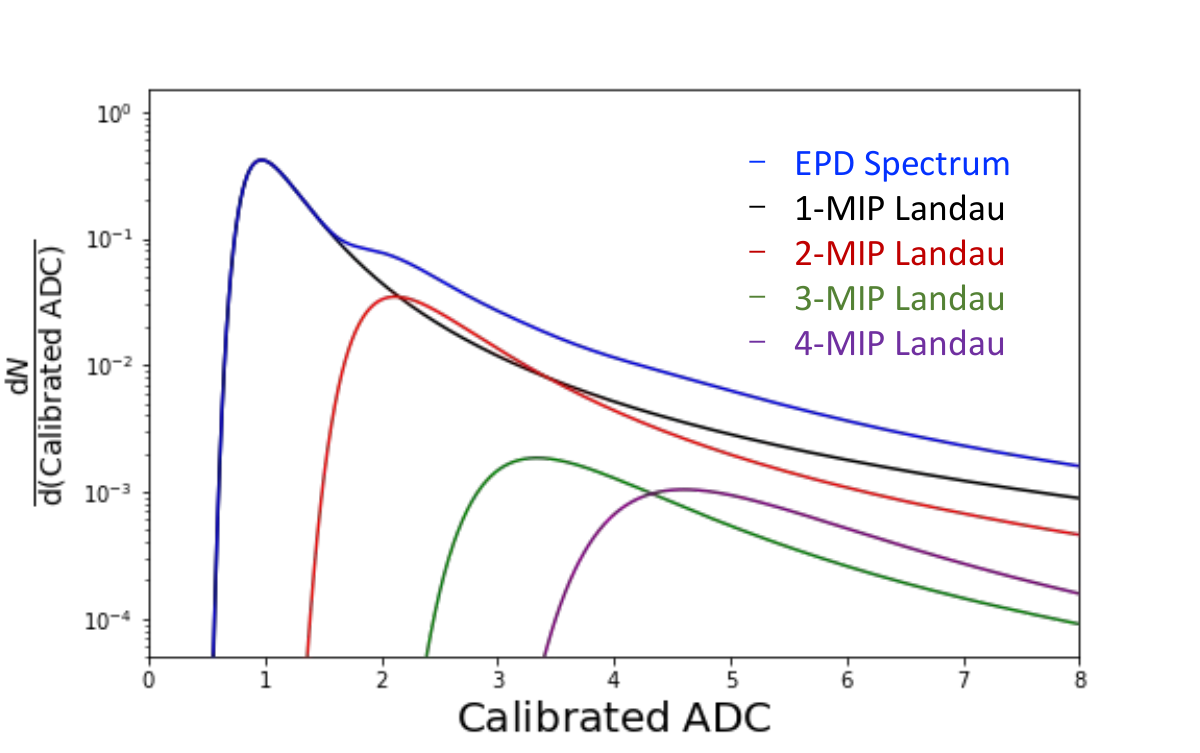}
    \caption{The black, red, green, purple curves correspond to the 1,2,3,4-MIP Landau respectively. The blue curve shows what the EPD spectrum is like when $30\%$ of the events are a 1-MIP event, $5\%$ of the events are a 2-MIP event, $0.4\%$ of the events are a 3-MIP event, $0.3\%$ of the events are a 4-MIP event and the rest are a 0-MIP event. This plot is just a sketch for the purpose of demonstration, it is not made from real data.}
    \label{fig:nmip_fit_demo_sum_new}
\end{figure}

The energy loss of MIPs in a scintillator follows a Landau distribution, the ADC spectra of the EPD tiles thus also follow Landau distributions. The Landau distribution only has two parameters: the most probable value (MPV) and the width over MPV (WID/MPV). In principle, the WID/MPV only depends on the material and thickness of the detector; and the ADC values are calibrated in a way that MPV is normalized to unity for the Landau distribution corresponding to a single MIP traversing the EPD tile in all the events. Figure~\ref{fig:nmip_fit_demo_sum_new} demonstrates how an ideal EPD spectrum should look when different numbers of MIPs traverse an EPD tile. 
In general, the n-MIP Landau is the convolution of (n-1)-MIP Landau with the 1-MIP Landau (n $\neq1$). 

In reality, an EPD tile gets hit by a varying number of MIPs in each event. The resulting EPD spectrum is thus a weighted sum of the 1, 2, 3, $\dots$, n-MIP Landaus, with the weights representing the probabilities of 1, 2, 3, $\dots$, n-MIP events. This spectrum is illustrated by 
the blue curve in Figure~\ref{fig:nmip_fit_demo_sum_new}. 
Therefore, the distribution of calibrated ADCs can be described by the equation:
\begin{equation}\label{eq_landau}
\frac{\text{d}N}{\text{d(Calibrated ADC)}}=\\
\sum_{i=1}^{n} M_i L_i(\text{Calibrated ADC}),
\end{equation}
where $M_i$ represents the probability of the i-MIP events and $L_i$ represents the i-MIP Landau distribution: ($*$ is the convolution product.)
\begin{equation}
L_i=
    \begin{cases}
      \text{Landau(MPV,WID/MPV)}, & \text{if}\ i=1; \\
      L_{i-1}*L_1, & \text{otherwise}.
    \end{cases}
\end{equation}

Since the mean of the Landau distribution is undefined, the law of large numbers doesn't apply. Therefore, the mean of calibrated ADCs will not necessarily get close to the averaged number of MIPs 
as more data gets collected. Instead, the probability of the i-MIP event must be derived by fitting the measured calibrated ADC spectra with Eq.~\ref{eq_landau}. Assuming the contribution to such spectra comes from up to 4-MIP events, such a fitting will involve six parameters $M_1,M_2,M_3$, $M_4$, MPV and WID/MPV. So, the averaged number of particles in each tile can be calculated as:\begin{equation}\label{eq:averageN}
     N=\sum_{i=1}^{4} M_{i}\times i.
\end{equation}
The corresponding uncertainty on $N$ needs to be calculated by the covariance matrix:
\begin{equation}\label{eq:averageNerr}
   \sigma^{2}=\mathbf{k}\mathbf{\Sigma}\mathbf{k}^{\top},
\end{equation}
where $\mathbf{\Sigma}$ is the covariance matrix of the fitting parameters and $\mathbf{k}=(1,2,3,4,0,0)$ is the partial derivatives of N over those fitting parameters. 

\begin{figure*}
    \centering
    \includegraphics[width=\textwidth]
    {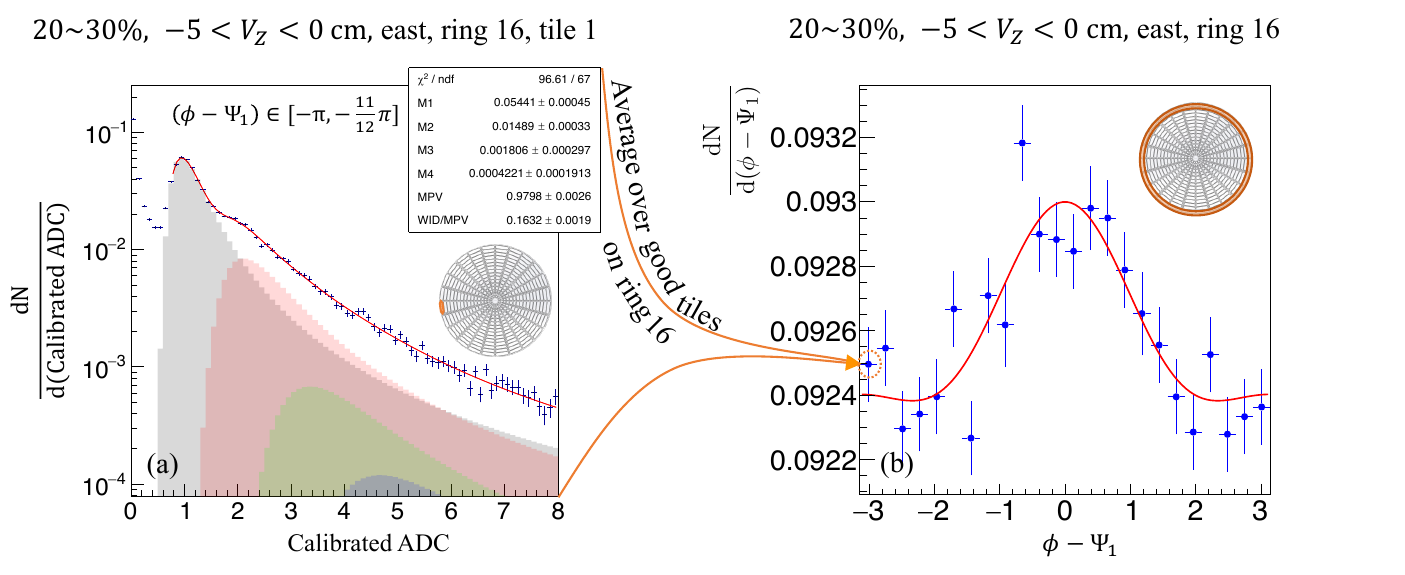}
    \caption{(a): the calibrated ADC spectra (blue, largely obscured by the red curve) for tile 1 at ring 16 on the east EPD (shown by the orange area on the middle right EPD sketch) obtained from events with $(\phi-\Psi_{1})\in[-\pi,-\frac{11}{12}\pi]$, $-5<V_z<$0 cm, and 20-30\% centrality. The shaded areas represent the expected calibrated ADC spectra when 1, 2, 3, 4 minimum ionization particles (MIPs) traverse an EPD tile. The histogram is fitted by the weighted sum of these four distributions (red curve) using the weights ($M_{i}$) as the fitting parameters. The fitting is applied for a calibrated ADC range from 0.75 to 8, to exclude background noise. (b): the $\mathrm{d}N/\mathrm{d}(\phi-\Psi_{1}^{\mathrm{TPC}})$ distribution for ring 16 obtained from all the events (shown by the orange area on the upper right EPD sketch). Each point is obtained by fitting multiple calibrated ADC spectra. The leftmost point is calculated from the calibrated ADC spectra of 24 tiles on ring 16 including the one shown in (a).}
    \label{fig:nmip_fit_data}
\end{figure*}

As illustrated in Figure~\ref{fig:epd_illustration}, the pseudorapidity coverage of each EPD tile depends on $V_{z}$. Therefore, this analysis was carried out in sixteen $V_{z}$ bins in $[-40,40]$ cm. The $\eta$ and $\phi$ of each EPD tile is determined by a straight line between the primary vertex and a uniformly distributed random point on the EPD tile. The widths of the EPD tiles are incorporated into the systematic uncertainties on $\eta$. Figure~\ref{fig:nmip_fit_data} demonstrates the procedure of extracting $v_{1}$ of Ring 16 on the east EPD for 20-30\% centrality and $-5<$\vz$<0$ cm. First of all, for each tile on Ring 16, we make the calibrated ADC spectrum for each (\deltaphi) bin and apply the fit. 
Figure~\ref{fig:nmip_fit_data} (a) is made from events in which the difference between the $\phi$ of Tile 1 and \psitpc\ of the event is between $-\pi$ and $-\frac{11}{12}\pi$. Then, the $\mathrm{d}N/\mathrm{d}(\phi-\Psi_{1}^{\mathrm{TPC}})$ distribution for a single tile can be calculated by Eq.~\ref{eq:averageN}. 
Next, we take the average of all the good tiles on Ring 16 and obtain the $\mathrm{d}N/\mathrm{d}(\phi-\Psi_{1}^{\mathrm{TPC}})$ distribution for the whole EPD ring as shown in Figure~\ref{fig:nmip_fit_data} (b).

\begin{figure}
    \centering
    \includegraphics[width=\columnwidth]
    {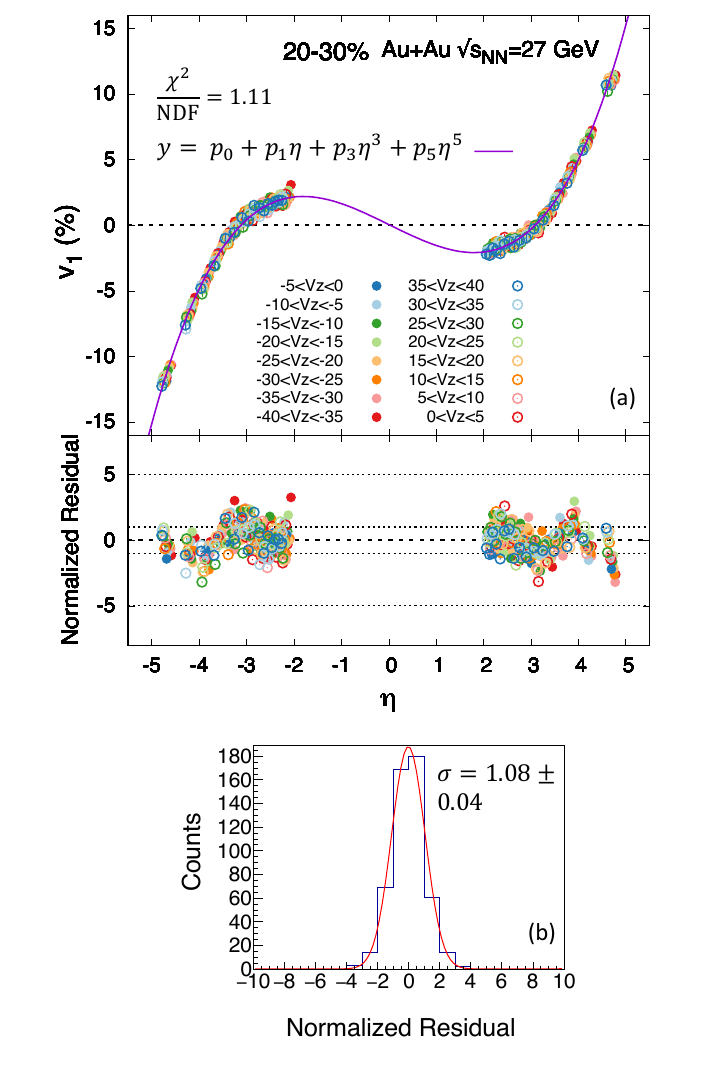}
    \caption{$v_{1}(\eta)$ for 16 $V_{z}$ bins between −40 cm and 40 cm, before correcting for the influence from the STAR material budget. All the data points are fitted by a smooth curve and the normalized residuals (residuals divided by error bars) follow a Gaussian distribution  of $\sigma\sim 1.08$, which indicates the fluctuations and error bars on the data points are reasonable. }
    \label{fig:v1_sanity_check}
\end{figure}

After obtaining the particle azimuthal distributions, we can extract the raw \vone\ by fitting them with Fourier expansions:

\begin{equation}\label{eq:v1fit}
    \begin{aligned}
    &\frac{\mathrm{d}N}{\mathrm{d}(\phi-\Psi_{1}^{\mathrm{TPC}})}=\\
&k[1+2v_{1}\cos(\phi-\Psi_{1}^{\mathrm{TPC}})+2v_{2}\cos(2\phi-2\Psi_{1}^{\mathrm{TPC}})],
    \end{aligned}
\end{equation}
where the higher orders are left out. Then, we correct the raw \vone\ with the event-plane resolution:
\begin{equation}\label{eq:v1_reso_correction}
	v_{1}^{\mathrm{measured}}=\frac{v_{1}}{R_1^{\mathrm{TPC}}}.
\end{equation}
Figure~\ref{fig:v1_sanity_check}(a) shows $v_{1}(\eta)$ after the resolution correction for sixteen \vz\ bins for 20-30\% centrality at \snn\ 27~GeV. The \vone\ obtained from Figure~\ref{fig:nmip_fit_data}(b) only corresponds to one data point in Figure~\ref{fig:v1_sanity_check}(a). As a sanity check, we fit all the data points with a smooth curve ($y(x)$) and calculate the normalized residuals (shown by the lower panel of Figure~\ref{fig:v1_sanity_check}(a)):
\begin{equation}\label{eq:normalized_residual}
r_{i}=\frac{y(x_{i})-f(x_{i})}{\sigma_{i}},
\end{equation}
where $f(x)$ is the fitting function and $\sigma_{i}$ is the error bar associated with the data point.
The normalized residuals follow a Gaussian distribution of $\sigma\sim1.08$ (Figure~\ref{fig:v1_sanity_check}(b)), which indicates that the fluctuations and error bars on the data points are reasonable. 

In order to better present the final results, we group every sixteen \vone\ points along $\eta$ simply by taking the average of the sixteen \vone\ values and sixteen $\eta$ values. Note that the group of sixteen points might have contributions from different EPD rings especially at small $|\eta|$. This underscores the importance of measuring \veta\ in small \vz\ bins instead of using a wide \vz\ range.

\subsection{\label{sec:level3_3}Correction for STAR material effect}
\begin{figure}
    \centering
    \includegraphics[width=1\columnwidth]{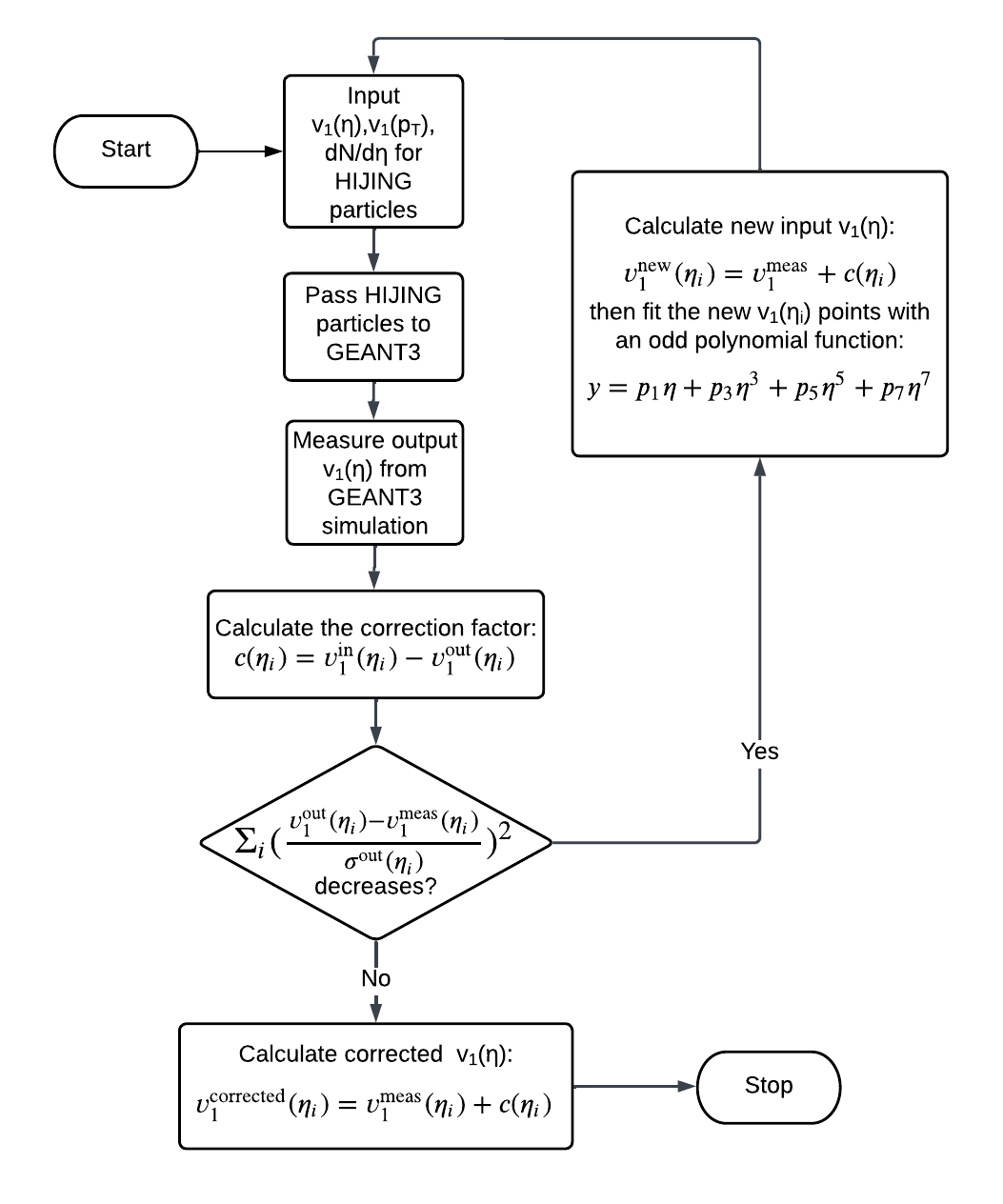}
    \caption{Flow chart for the \texttt{HIJING+GEANT3} correction.}
    \label{fig:geant4_flow_chart_ver2}
\end{figure}

Simulation showed that about half of the particles the EPD detects are produced via interactions between the primary particles and the materials in the detector. 
In order to correct for this material effect, we carried out an iterative process as shown by the flow chart in Figure~\ref{fig:geant4_flow_chart_ver2}. Monte Carlo events were generated by HIJING~\cite{Wang:1991hta} and passed to the GEANT3~\cite{Brun:1994aa} simulation of the STAR detector. The HIJING model does not emit isotopes with $A>1$. The yield (\dndeta) and azimuthal distribution (\vpt\ and \veta) of the input HIJING tracks can be tuned to any desired shape by weighting. The same weight needs to be applied simultaneously to the HIJING track and its daughters produced in the GEANT3 simulation. Then, we use the difference between the input and output \veta\ (i.e correction factor $c(\eta)$) to create a new input \veta\ and repeat the GEANT3 simulation. 

The goal of this iterative process is to find the input \veta\ that can reproduce the measured \veta, so that the corresponding $c(\eta)$ can be used to calculate the \veta\ of the primary particles:
\begin{equation}
v_1^{\text{corrected}}(\eta) = v_1^{\text{measured}}(\eta)+c(\eta)
\end{equation}
The iteration is terminated when the difference between the output and measured \veta\ stops decreasing. Note that the input \vone\ and \dndeta\ need to be continuous functions in the entire $\eta$ range ($|\eta|<6$) since primary particles out of the EPD
acceptance can also result in EPD hits through decaying and scattering. This process was conducted centrality by centrality and $V_z$-bin by $V_z$-bin. The correction due to the STAR material effect ($\frac{|c(\eta)|}{v_{1}^{\text{measured}}(\eta)}$) ranges from $\sim 1\%$ to $\sim100\%$. 

\section{\label{sec:level4}Systematic uncertainty}
When we correct for the STAR material effect with the HIJING+GEANT3 simulation, \dndeta\ and \vpt\ are required as input parameters. However, they are unknown simply because our detector doesn't cover the whole $\eta$ range and \pt\ cannot be measured at forward and backward $\eta$. Therefore, we took our best guess as the default setting, then vary \dndeta\ and \vpt\ within a reasonable range and incorporate the differences into systematic uncertainties. In the default setting, \vone\ is independent of \pt.  
The variations are as follows, only one parameter is varied at a time: 
\begin{enumerate}
	\item $v_1(p_{T})=k\cdot \sqrt{p_{T}}$;
	\item $v_1(p_{T})=k\cdot p_{T}^2$;
	\item \dndeta. 
\end{enumerate}

\begin{figure*}
    \hspace*{-2cm}
    \includegraphics[scale=0.5]{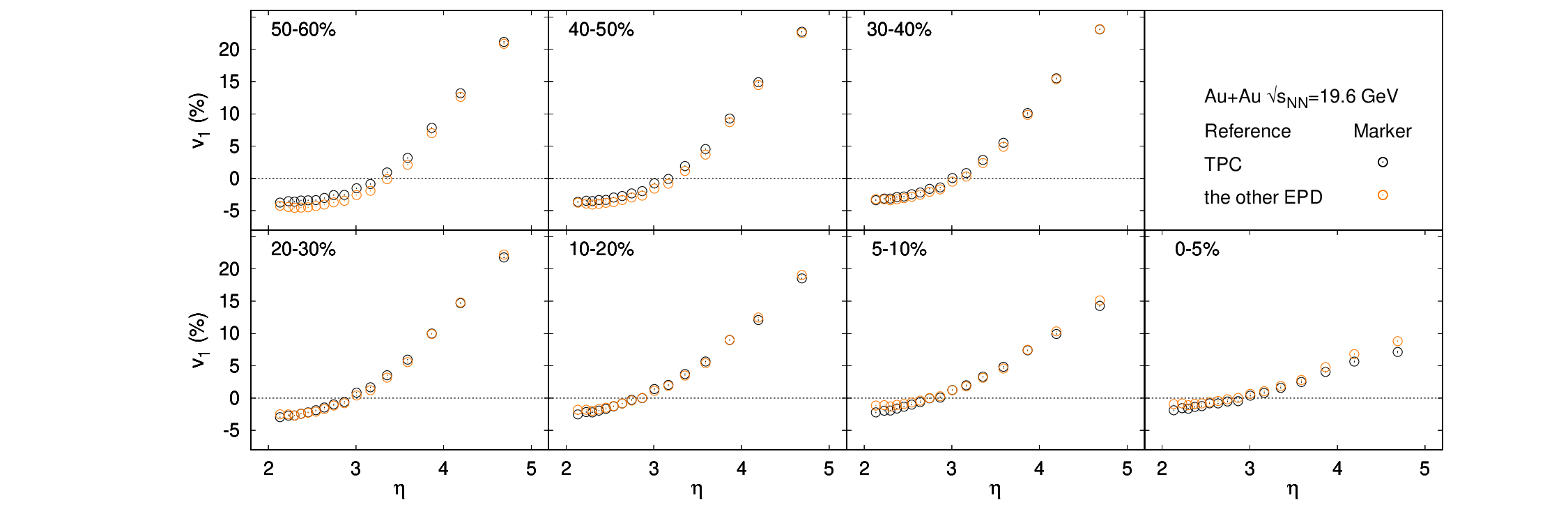}
    \caption[\veta\ measured with two different references at \snn\ 19.6 GeV]{The \veta\ measured with two different references at \snn\ 19.6 GeV (before correcting for the STAR material budget). Orange data points are measured with respect to \psione\ from the other side of the EPD; black data points are measured with respect to \psione\ from the TPC. 
The results from the forward and backward pseudorapidities are combined by averaging \veta\ and  $-v_1(-\eta)$. 
The statistical uncertainties on \vone\ and $\eta$ are plotted as vertical and horizontal lines at the center of the markers respectively. Some of them are too small to be visible.}
    \label{fig:Figure_v1_poseta_otherEPD_19p6}
\end{figure*}
\begin{figure*}
    \hspace*{-2cm}
    \includegraphics[scale=0.5]{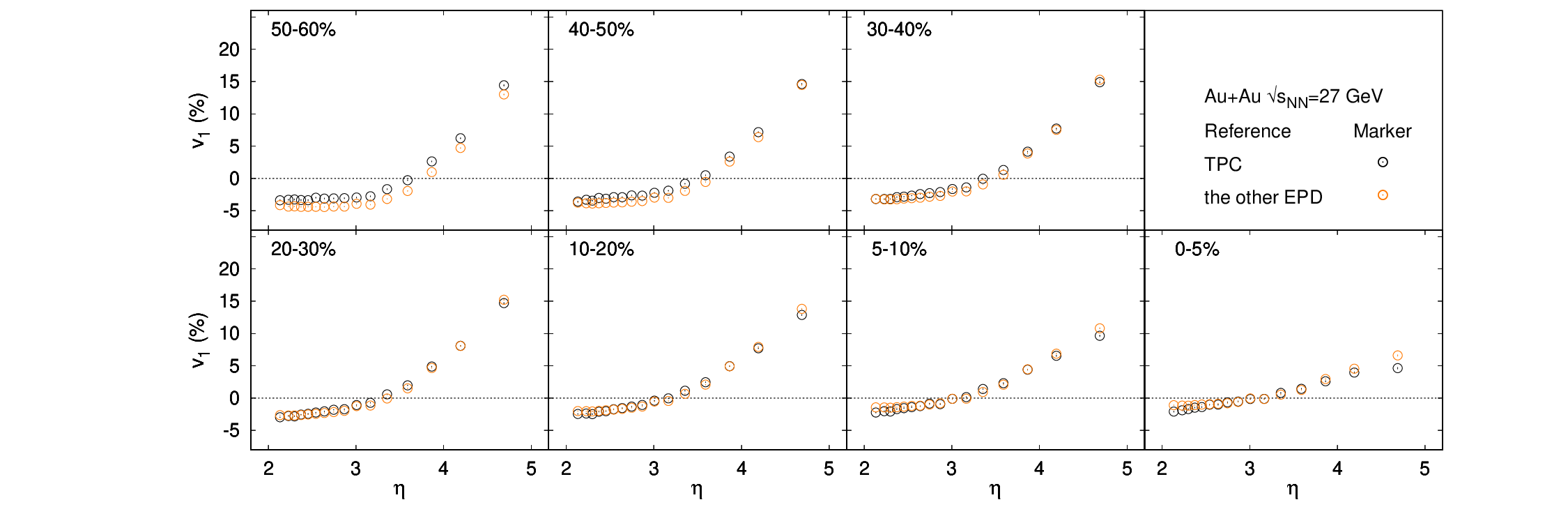}
    \caption[\veta\ measured with two different references at \snn\ 27 GeV]{The \veta\ measured with two different references at \snn\ 27 GeV (before correcting for the STAR material budget), similar to Figure~\ref{fig:Figure_v1_poseta_otherEPD_19p6}.}
    \label{fig:Figure_v1_poseta_otherEPD_27}
\end{figure*}

\begin{figure*}
    \hspace*{-2cm}
    \includegraphics[scale=0.5]{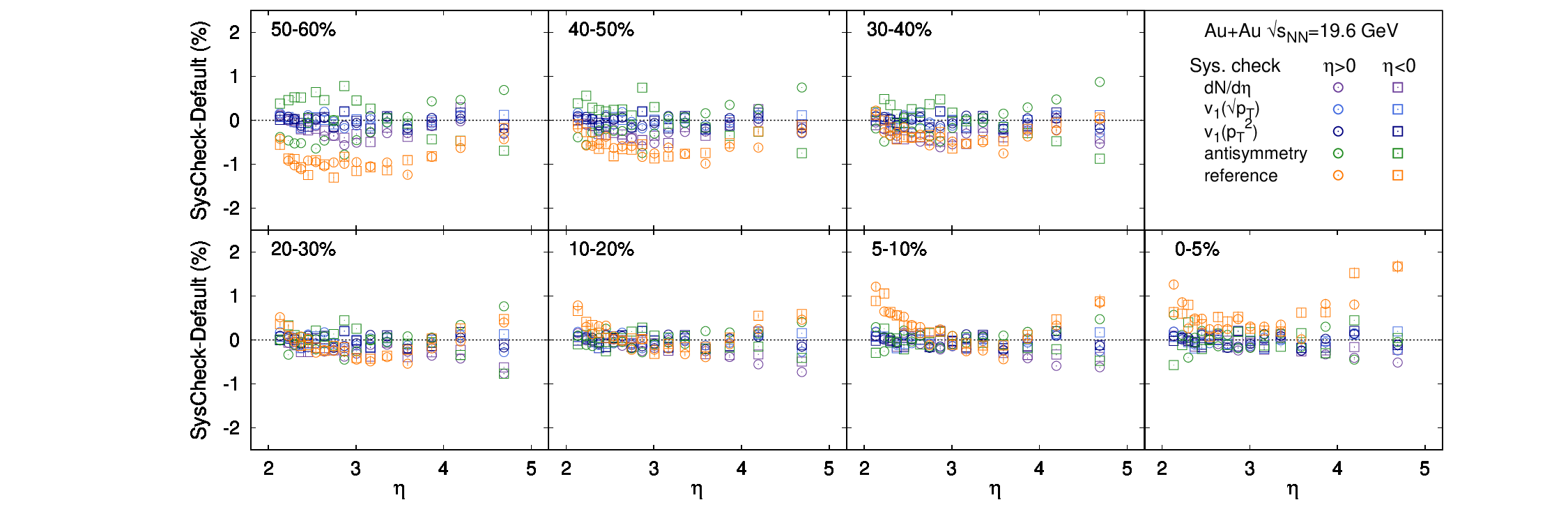}
    \caption{Multiple systematic checks at \snn\ 19.6 GeV. Different colors represent different checks. Circles are $y(\eta)$ at $\eta>0$, squares are  $-y(-\eta)$ at $\eta<0$. Refer to the text for the formulas used to calculate the value and error bar of each point. Only the points whose error bars do not touch zero are incorporated into the systematic uncertainties.}
    \label{fig:Figure_v1_Sys_diff_19p6}
\end{figure*}

\begin{figure*}
    \hspace*{-2cm}
    \includegraphics[scale=0.5]{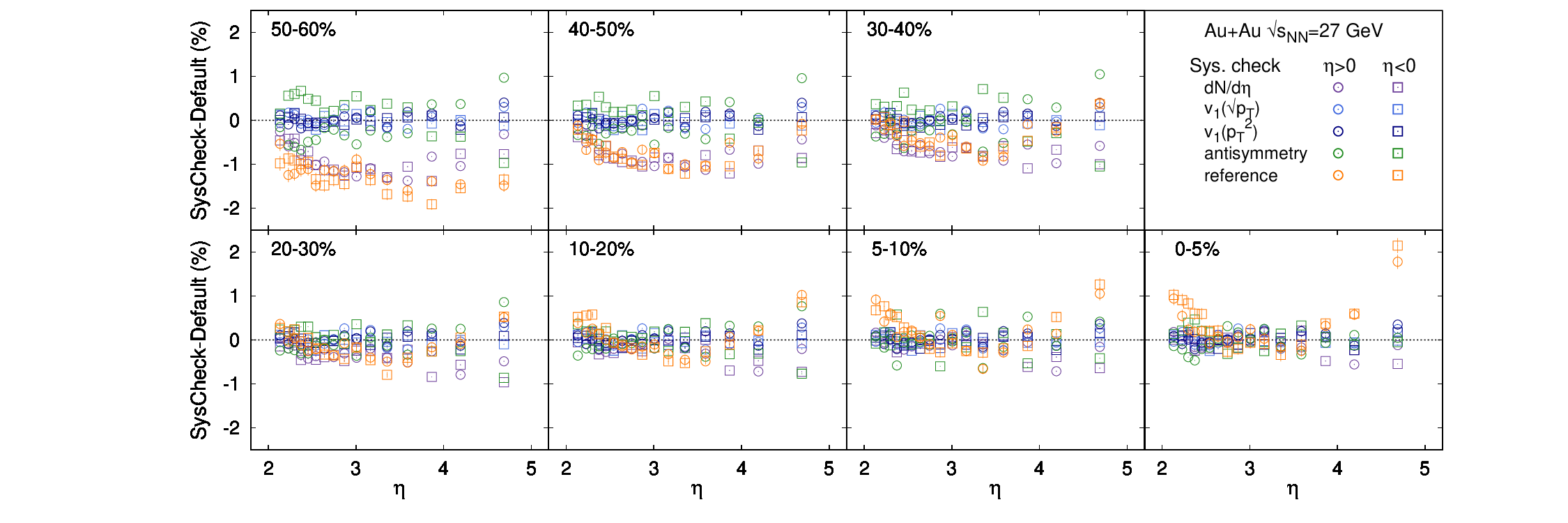}
    \caption{Multiple systematic checks at \snn\ 27 GeV, similar to Figure~\ref{fig:Figure_v1_Sys_diff_19p6}.}
    \label{fig:Figure_v1_Sys_diff_27}
\end{figure*}

The directed flow was also measured with respect to the event plane reconstructed by the other side of the EPD as a systematic check. The East (West) EPD resolution is also calculated by the ``three sub-event method":
\begin{equation}\label{eq:tpc_resolution}
    \begin{aligned}
    &R_{1}^{\mathrm{EPDE(W)}}=\\
&\sqrt{\frac{\langle\cos{(\Psi_{1}^{\mathrm{TPC}}-\Psi_{1}^{\mathrm{EPDE(W)}})}\rangle\langle\cos{(\Psi_{1}^\mathrm{EPDE}-\Psi_{1}^\mathrm{EPDW})}\rangle}{\langle\cos{(\Psi_{1}^\mathrm{TPC}-\Psi_{1}^\mathrm{EPDW(E)})}\rangle}}.
    \end{aligned}
\end{equation}
Figures~\ref{fig:Figure_v1_poseta_otherEPD_19p6} and~\ref{fig:Figure_v1_poseta_otherEPD_27} show the \veta\ measured with different references. The black data points are obtained using $\Psi_{1}$ from the TPC, while the orange data points are obtained using $\Psi_{1}$ from the other side of the EPD. 
Clear differences were observed especially at peripheral and central collisions. Further investigation shows the discrepancy could arise from three sources:

\begin{itemize}
\item Short-range correlations: the $\eta$ gap between the TPC and EPD is approximately equal to one for the outermost EPD ring, while the $\eta$ gap is at least four between the two EPD wheels. Non-flow effects like resonance decays and minijet fragmentation usually contribute to positive azimuthal correlations between two nearby $\eta$ regions. A larger $\eta$ gap will better suppress these short-range correlations, resulting in a smaller measured $|v_1|$. Therefore, $|v_{1}\{\Psi_{1}^{\mathrm{TPC}}\}|$ is expected to be larger than $|v_{1}\{\Psi_{1}^{\mathrm{EPD}}\}|$ especially at small $|\eta|$, as the PoI is closer to the reference. 
\item Momentum-conservation effect: when a single side of the EPD is used as the reference, the momentum-conservation effect cannot be suppressed, and it is more prominent when $\langle p_{T} \rangle$ is high (usually at small $|\eta|$) and the multiplicity is low (peripheral collisions). 
The momentum conservation effect has been clearly observed around midrapidity. While \vone\ at midrapidity should be zero simply due to the asymmetry of the collision, it was measured to be greater than zero when using $\Psi_{1}$ from the West EPD ($\eta>0$), and smaller than zero when using $\Psi_{1}$ from the East EPD ($\eta<0$) in all centralities. The \veta\ curve becomes an odd function when both sides of the EPD are used to reconstruct $\Psi_{1}$. The UrQMD (Ultra-relativistic Quantum Molecular Dynamics) simulation also yielded a ``\veta\ shift" with the same order of magnitude and sign. In this systematic check, at peripheral collisions, \veta\ measured with $\Psi_{1}$ from the East (West) EPD shifted down (up) compared to $|v_{1}\{\Psi_{1}^{\mathrm{TPC}}\}|$, and the shift is more prominent at smaller $|\eta|$. It is consistent with the expectation from the momentum-conservation effect.  
\item First-order event-plane decorrelation: at large $|\eta|$, both short-range correlations and the momentum-conservation effect will lead to $|v_{1}\{\Psi_{1}^{\mathrm{TPC}}\}|>|v_{1}\{\Psi_{1}^{\mathrm{EPD}}\}|$. However, $|v_{1}\{\Psi_{1}^{\mathrm{TPC}}\}|<|v_{1}\{\Psi_{1}^{\mathrm{EPD}}\}|$ was observed in central collisions, which might be explained by the first-order event-plane decorrelation. Model studies have shown that the first-order participant plane and the first-order spectator plane can be decorrelated due to the conservation of angular momentum. This decorrelation is most prominent at central and peripheral collisions~\cite{Adams:2021yob}. As a result, at smaller $|\eta|$, $|v_{1}\{\Psi_{1}^{\mathrm{TPC}}\}|$ can be larger than $|v_{1}\{\Psi_{1}^{\mathrm{EPD}}\}|$ because the PoI are more correlated with the participant plane (approximated by $\Psi_{1}^{\mathrm{TPC}}$); at larger $|\eta|$, $|v_{1}\{\Psi_{1}^{\mathrm{TPC}}\}|$ can be smaller than $|v_{1}\{\Psi_{1}^{\mathrm{EPD}}\}|$ because the PoI are more correlated with the spectator plane (approximated by $\Psi_{1}^{\mathrm{EPD}}$). As shown in Figures~\ref{fig:Figure_v1_Sys_diff_19p6} and~\ref{fig:Figure_v1_Sys_diff_27}, the possible event-plane decorrelation signals are beyond other systematic uncertainties in 0-5\% centrality.
\end{itemize}
Note the above three effects can influence the measurement of event plane resolutions, too. Ideally, only the short-range correlations should be included in the systematic uncertainty. However, it is impractical to isolate and disentangle the individual effects of short-range correlations. Therefore, the differences between $|v_{1}\{\Psi_{1}^{\mathrm{TPC}}\}|$ and $|v_{1}\{\Psi_{1}^{\mathrm{EPD}}\}|$ are directly incorporated into the systematic uncertainties. 

Lastly, \veta\ should be an odd function due to the asymmetry of the collision. Thus, the consistency between $|v_{1}(\eta)|$ at forward and backward $\eta$ becomes a natural systematic check. 

The total systematic uncertainty was calculated using Barlow's method~\cite{Barlow:2002yb}. For the $i^{\mathrm{th}}$ systematic check, calculate:
\begin{equation}\label{eq:y_diff}
Y_{\mathrm{diff}}=Y_i-Y_{\mathrm{df}},
\end{equation}
\begin{equation}\label{eq:sigma_diff}
\sigma_{\mathrm{diff},i}=\sqrt{|\sigma^2_{\mathrm{stat},i}-\sigma^2_{\mathrm{stat,df}}|},
\end{equation}
where $Y_{\mathrm{df}}$ is the observable measured in the default analysis, $\sigma_{\mathrm{stat,df}}$ is the associated statistical uncertainty; $Y_{i}$ is the observable measured in the systematic check, and $\sigma_{\mathrm{stat},i}$ is the associated statistical uncertainty. Then:
\begin{equation}
    \sigma_{\mathrm{sys},i}= \begin{cases}
0, &Y_{\mathrm{diff}}\le\sigma_{\mathrm{diff},i},\\
\sqrt{Y_{\mathrm{diff}}^2-\sigma_{\mathrm{diff},i}^2}, &Y_{\mathrm{diff}}>\sigma_{\mathrm{diff},i}.
\end{cases}
\end{equation}
The final systematic uncertainty is calculated as:
\begin{equation}
 \begin{aligned}
    &\sigma_{\mathrm{sys}}=\\
&\sqrt{(\frac{\sigma_{\mathrm{sys},1}}{\sqrt{3}})^2+(\frac{\max(\sigma_{\mathrm{sys},2},\sigma_{\mathrm{sys},3})}{\sqrt{3}})^2+\sum_{i=4}^{5}(\frac{\sigma_{\mathrm{sys},i}}{\sqrt{12}})^2},
    \end{aligned}
\end{equation}
where systematic check 1 is the variation of input \dndeta\ in the GEANT correction; systematic checks 2 and 3 are the variations of \pt\ dependence of \vone; systematic check 4 is the $|v_{1}|$ difference between $\eta>0$ and $\eta<0$; systematic check 5 is the variation of reference. Figures~\ref{fig:Figure_v1_Sys_diff_19p6} and~\ref{fig:Figure_v1_Sys_diff_27} show the results for all the systematic checks. The data points were calculated by Eq.~\ref{eq:y_diff} and the error bars were calculated by Eq.~\ref{eq:sigma_diff}. Only the points which are non-zero within the error bars were incorporated into the systematic uncertainties.

\section{\label{sec:level5}Results}
\begin{figure*}
    \hspace*{-3cm}
    \includegraphics[scale=0.6]{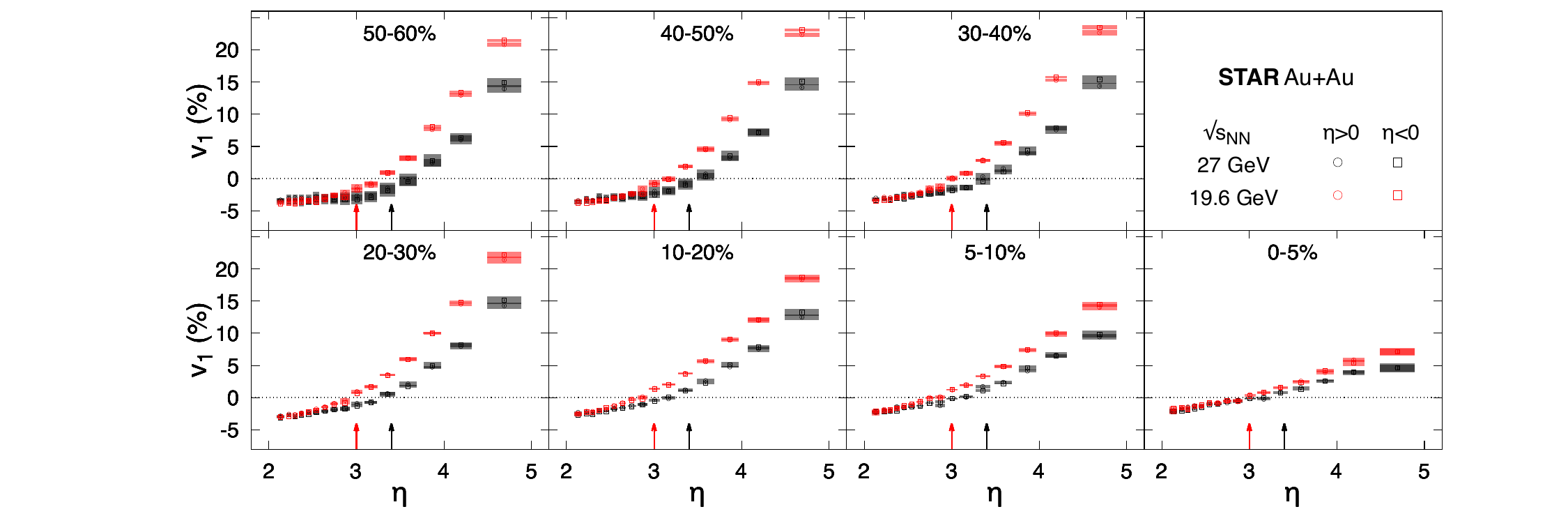}
    \caption[\veta\ at \snn19.6 and 27 GeV]{\veta\ at \snn\ 19.6 (red) and 27 (black) GeV in Au+Au collisions for seven centralities. Circles are $v_{1}(\eta)$ at $\eta>0$, squares are  $-v_{1}(-\eta)$ at $\eta<0$. The statistical uncertainties on \vone\ and $\eta$ are plotted with vertical and horizontal lines respectively. The statistical uncertainties associated with $\eta$ are too small and hidden behind the line widths. The systematic uncertainties are plotted with boxes whose heights and widths represent the systematic uncertainties on \vone\ and $\eta$ respectively. The red arrows represent the beam rapidity ($y_{\mathrm{beam}}=3.0$) at \snn19.6 GeV, while the black arrows represent the beam rapidity ($y_{\mathrm{beam}}=3.4$) at \snn\ 27 GeV.}
    \label{fig:Figure_v1_poseta_xsys_new}
\end{figure*}

Figure~\ref{fig:Figure_v1_poseta_xsys_new} shows \veta\ measured in Au+Au collisions at \snn\ 19.6 and 27 GeV for seven centralities. 
At both energies, \veta\ crosses zero roughly around the beam rapidity for all the centralities. 
At $\eta<y_{\mathrm{beam}}$, $|v_{1}(\eta)|$ decreases towards central collisions; at $\eta>y_{\mathrm{beam}}$, $|v_{1}(\eta)|$ slightly increases and then decreases going from peripheral to central collisions.

The ``\vone\ wiggle" (\vone\ changes sign three times along $\eta$ including the zero crossing at midrapidity) has been observed by multiple experiments at various energies~\cite{PHOBOS:2005ylx, STAR:2005btp, STAR:2003xyj, STAR:2008jgm, STAR:2011gzz}. It is believed to be due to the longitudinal hydrodynamic expansion of a tilted source~\cite{Bozek:2010bi, Chatterjee:2017ahy, Bozek:2022svy, PhysRevC.105.034901,Du:2022yok,Nara:2021fuu}. A tilted source is created after the collision due to the local imbalance of the longitudinal momenta of the forward- and backward-going participants. Due to the higher pressure gradient, more particles are produced along the minor axis of the tilted fireball, leading to a negative \veta\ slope around midrapidity. For central collisions, the fireball is less tilted and less anisotropic, resulting in a milder \veta\ slope at small $|\eta|$~\cite{Bozek:2010bi}. For peripheral collisions, the fireball is more titled, but the nuclear fragments also receive a stronger deflection, leading to both large flow and large ``anti-flow" in the fragmentation region. This could explain the nonmonotonic change of \vone\ with centrality at forward $\eta$. 

\begin{figure*}
    \hspace*{-3cm}
    \includegraphics[scale=0.6]{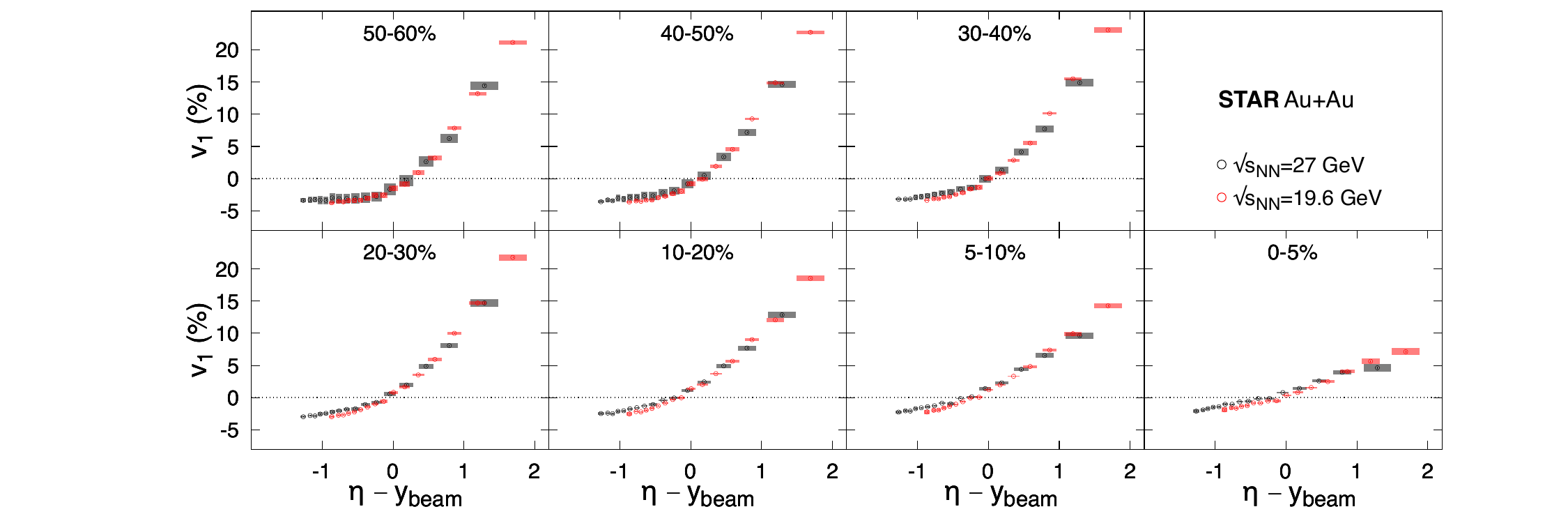}
    \caption[$v_1(\eta-y_{\mathrm{beam}})$ at \snn19.6 and 27 GeV]{$v_1(\eta-y_{\mathrm{beam}})$ at \snn\ 19.6 (red) and 27 (black) GeV in Au+Au collisions for seven centralities. The results from the forward and backward pseudorapidities are combined by averaging \veta\ and  $-v_1(-\eta)$. Statistical uncertainties are plotted with lines, systematic uncertainties are plotted with boxes.}
    \label{fig:Figure_v1_poseta_LF_xsys_new}
\end{figure*}

Figure~\ref{fig:Figure_v1_poseta_LF_xsys_new} shows $v_{1}(\eta-y_{\mathrm{beam}})$ at \snn\ 19.6 and 27 GeV for seven centralities. The results from the forward and backward pseudorapidities are combined by averaging \veta\ and  $-v_1(-\eta)$. 
At all centralities, the $v_{1}(\eta-y_{\mathrm{beam}})$ curves from two energies coincide within uncertainties especially beyond the beam rapidity. 

\begin{figure}
    \hspace*{-1cm}
    \includegraphics[width=1.2\columnwidth]{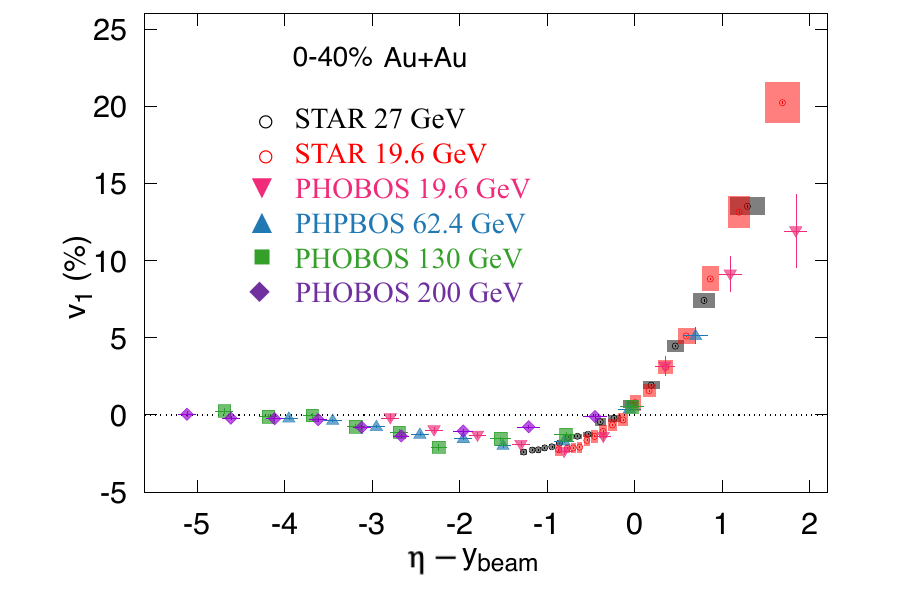}
    \caption[STAR $v_{1}(\eta-y_{\mathrm{beam}})$ compared with PHOBOS]{$v_{1}(\eta-y_{\mathrm{beam}})$ measured by STAR and PHOBOS for $0\sim40\%$ centrality. Only statistical uncertainties are plotted for the PHOBOS results, while both statistical uncertainties (lines) and systematic uncertainties (boxes) are plotted for the STAR results.}
    \label{fig:v1_phobos}
\end{figure}

This energy scaling with $(\eta-y_{\mathrm{beam}})$ is usually referred as ``limiting fragmentation." The hypothesis of limiting fragmentation states that, in high-energy collisions, two incoming particles go through each other and break into fragments in the process instead of completely stopping each other~\cite{Benecke:1969sh}. It further predicts that at sufficiently-high energies, both $\mathrm{d}^2N/\mathrm{d}y'\mathrm{d}p_{T}$ and the mix of particles species reach a limiting value and become independent of energy in a region around $y'\sim0$, where $y'\equiv y-y_{\mathrm{beam}}$ and $y$ is the rapidity. It also implies a limiting value for $\mathrm{d}N/\mathrm{d}\eta'$ where $\eta'\equiv\eta-y_{\mathrm{beam}}$~\cite{Back:2002wb}. This $\mathrm{d}N/\mathrm{d}\eta'$ scaling has been observed both at BRAHMS and PHOBOS~\cite{BRAHMS:2001gci,BRAHMS:2004gnf,PHOBOS:2010eyu}. Surprisingly, the same scaling behavior was also observed for directed and elliptic flow at PHOBOS and STAR ~\cite{PHOBOS:2004nvy,PHOBOS:2005ylx,STAR:2005btp,STAR:2003xyj,STAR:2008jgm,STAR:2011gzz}. This analysis verified the $v_{1}(\eta')$ scaling again at one more energy with high precision (Figure \ref{fig:v1_phobos}). While the energy scaling of the yield around $\eta'\sim0$ can be attributed to ``spectators" minimally influenced by the collisions, the energy scaling of directed flow is less intuitive to comprehend, as \vone\ is usually closely related to the collision dynamics. A common interpretation for large \vone\ at the forward rapidity is the deflection of nuclear fragments. However, it is hard to explain the energy independence of the directed flow around $\eta'\sim0$ with this picture. The limiting fragmentation of directed flow indicates the production of \vone\ at the fragmentation region might not only come from the deflection. Figure~\ref{fig:v1_phobos} also shows the comparison between the STAR and PHOBOS measurements. The results at \snn\ 19.6 GeV exhibit excellent consistency.

\begin{figure}[h]
    \centering
    \includegraphics[width=\columnwidth]{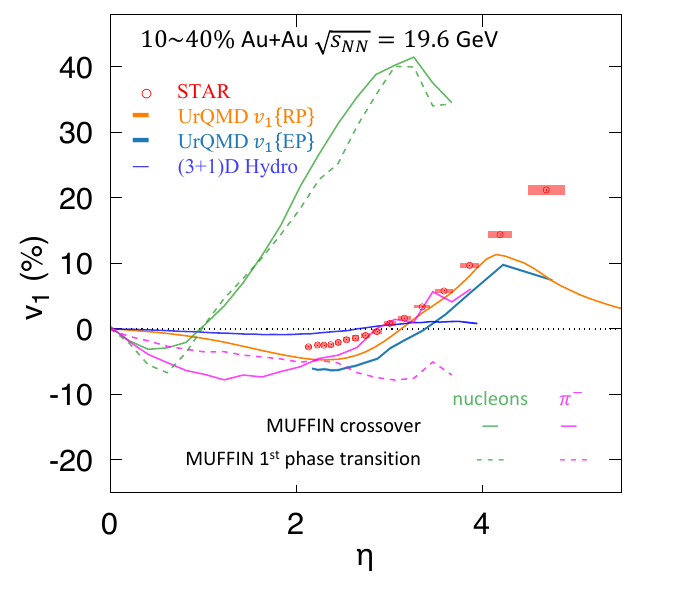}
    \caption[Model comparison at \snn19.6 GeV]{Model comparisons of \veta\ at \snn\ 19.6 GeV for $10\sim40\%$ centrality. Details about the models can be found in the text. Note \vone\ from the MUFFIN simulations are measured as a function of rapidity instead of the pseudorapidity. Statistical uncertainties are plotted as bands for the UrQMD and (3+1)D Hydro results, but sometimes they are too small and hidden behind the line width.}
    \label{fig:v1_19p6_model_cent1040}
\end{figure}

\begin{figure}[h]
    \centering
    \includegraphics[width=\columnwidth]{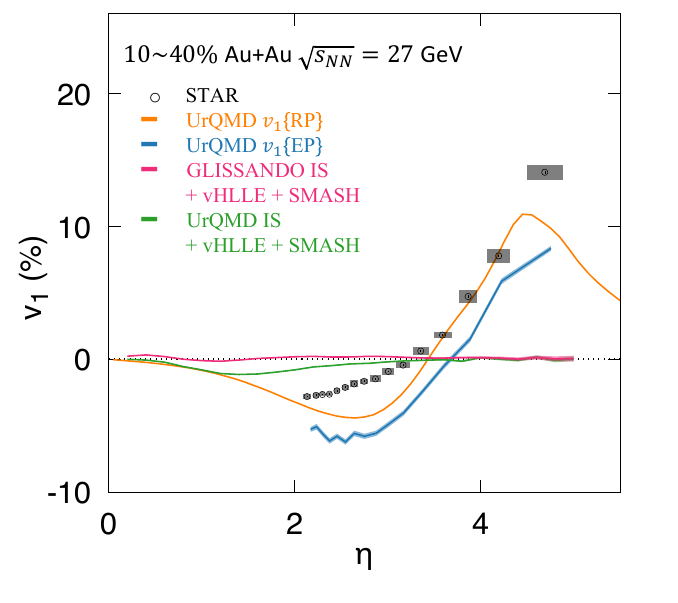}
    \caption[Model comparison at \snn27 GeV]{Model comparisons of \veta\ at \snn\ 27 GeV for $10\sim40\%$ centrality. Statistical uncertainties are plotted as bands for the UrQMD, but sometimes they are too small and hidden behind the line width.}
    \label{fig:v1_27_model_cent1040}
\end{figure}
Figure \ref{fig:v1_19p6_model_cent1040}  shows the model comparison at \snn\ 19.6 GeV. 
 UrQMD is a microscopic hadron transport and string model. In the standard cascade mode, UrQMD simulates the production of particles via hadron rescattering, resonance decays, and string excitation and decay. It doesn't rely on any mean-field or equilibrium assumptions. 
The UrQMD particles are sampled at 500 fm/$c$ after the collision in the cascade mode. 
The UrQMD \vone\ are calculated with respect to the reaction plane (the plane spanned by the impact parameter and the beam direction) and the event plane respectively. The event-plane angle and its resolution are calculated with exactly the same formula, reference, track cuts and weights as what were used in the experimental measurement. The discrepancy between \vone\{RP\} and \vone\{EP\} can originate from the lumpiness of the colliding nuclei, the non-flow correlations and the decorrelation between the spectator plane and the participant plane. 
This underscores the importance of using the same reference when comparing the model studies and experimental measurements. 
Although UrQMD failed to reproduce the experiment results quantitatively, it was able to reproduce the overall shape of the data including the ``\vone\ wiggle" and the substantial nonzero \vone\ at forward $\eta$.

MUFFIN~\cite{Cimerman:2023hjw} is an event-by-event three-fluid dynamic model based on the vHLLE code~\cite{Karpenko:2013wva}. In this model, the incoming nuclei are represented by two droplets of cold nuclear fluid, called projectile and target fluids. The process of a heavy-ion collision is thus modeled as mutual interpenetration of the projectile and target fluids. The phenomenon of baryon stopping is modeled as the friction between the projectile and target fluids. The kinetic energy lost to friction is channeled into the creation of a third fluid, which represents particles produced in the reaction. In this calculation, MUFFIN was coupled to a final-state hadronic cascade using SMASH~\cite{Weil:2016zrk}. The \vone\ from MUFFIN are measured with respect to the reaction plane and as a function of rapidity. A \pt\ cut of $0.15<p_{T}<2.0$ GeV/$c$ was applied to the simulation data while no \pt\ cut was applied to the particles-of-interest region in the experiment. 
This MUFFIN+SMASH hybrid simulation shows the sign of $\pi^{-}$ \vone\ at large rapidity is sensitive to the QGP phase transition. It will be interesting to see if this sensitivity still exists for charged-particle \veta. Since the proportion of nucleons increases at the fragmentation region, it is possible that \vone\ at large $\eta$ is predominantly influenced by the nucleon \vone, which exhibits a mild dependence on the QGP phase transition.

A (3+1)-dimensional hybrid framework with parametric initial conditions (both the initial energy-density distribution and the initial baryon-density distribution) has been developed recently~\cite{Du:2022yok}. This model has successfully reproduced the measured rapidity (around the midrapidity) and beam-energy dependence of the directed flow $v_{1}(y)$ of identified particles from \snn\ 7.7 to 200 GeV. However, it yields significantly smaller \veta\ compared to the STAR measurement at the forward $\eta$. This discrepancy mainly arises from the fact that this model only takes into account the fluid at the participant region. In reality, the nucleons that do not directly overlap with other nucleons in the initial stage of the collision can also interact with the fireball, thus making a substantial contribution to the final particle production across the entire (pseudo)rapidity range. Therefore, a satisfactory understanding of heavy-ion collisions at the BES energies will require modeling the dynamics over the entire $\eta$ range. 

Figure \ref{fig:v1_27_model_cent1040} shows the model comparisons at \snn\ 27 GeV. Similar to \snn\ 19.6 GeV, UrQMD qualitatively reproduced the overall shape of the measured \veta. 
A one-fluid model was also utilized to compute the \vone, employing two different initial states (generated by GLISSANDO~\cite{Rybczynski:2013yba} and UrQMD respectively). This model failed to produce any non-zero \vone\ at the forward $\eta$. Again, this failure demonstrates the importance of including all the segments of the heavy-ion collisions in the model study.

\section{\label{sec:level6}Summary}
In this paper, we presented \veta\ measured in Au+Au collisions at \snn\ 19.6 and 27 GeV over six units of $\eta$ using the STAR EPD. 
In order to measure directed flow without reconstructing individual charged tracks, we developed an entirely-new method to ensure the accuracy of this measurement. It includes fitting the EPD spectra with convoluted Landau distributions and correcting for the STAR material effect with the GEANT3 simulation. The results at \snn\ 19.6 GeV exhibit excellent consistency with the previous PHOBOS measurement in the same phase space (\vone\ integrated over all $p_{T}$), while elevating the precision to a new level. The increased precision of the measurement also revealed finer structures of the heavy-ion collision, including potential evidence for the first-order event-plane decorrelation. A collision-energy scaling of $v_{1}(\eta-y_{\mathrm{beam}})$ was observed at  $(\eta-y_{\mathrm{beam}})>0$ for \snn\ 19.6 and 27 GeV, which indicates the \vone\ at large $|\eta|$ might not only come from the deflection of nuclear fragments. 

Simulations from various models including transport, hydrodynamic, one-fluid hybrid and three-fluid hybrid models have been compared to this measurement. Only UrQMD (transport model) and MUFFIN (three-fluid hybrid model) were able to reproduce a significant \vone\ at the forward(backward) $\eta$ as observed in the experiment. This underscores the importance of incorporating all segments of the heavy-ion collision in model studies, especially at BES energies where nuclear fragments can substantially influence particle production across the entire pseudorapidity range. The comparison with the MUFFIN simulation indicates \vone\ at large $\eta$ might be sensitive to the QGP phase transition. Furthermore, the UrQMD study has shown significant discrepancy between $v_{1}\{\mathrm{EP}\}$ and $v_{1}\{\mathrm{RP}\}$, demonstrating the importance of employing the same reference when comparing experimental measurements and model calculations. 

\begin{acknowledgments}
We thank the RHIC Operations Group and RCF at BNL, the NERSC Center at LBNL, and the Open Science Grid consortium for providing resources and support.  This work was supported in part by the Office of Nuclear Physics within the U.S. DOE Office of Science, the U.S. National Science Foundation, National Natural Science Foundation of China, Chinese Academy of Science, the Ministry of Science and Technology of China and the Chinese Ministry of Education, the Higher Education Sprout Project by Ministry of Education at NCKU, the National Research Foundation of Korea, Czech Science Foundation and Ministry of Education, Youth and Sports of the Czech Republic, Hungarian National Research, Development and Innovation Office, New National Excellency Programme of the Hungarian Ministry of Human Capacities, Department of Atomic Energy and Department of Science and Technology of the Government of India, the National Science Centre and WUT ID-UB of Poland, the Ministry of Science, Education and Sports of the Republic of Croatia, German Bundesministerium f\"ur Bildung, Wissenschaft, Forschung and Technologie (BMBF), Helmholtz Association, Ministry of Education, Culture, Sports, Science, and Technology (MEXT), Japan Society for the Promotion of Science (JSPS) and Agencia Nacional de Investigaci\'on y Desarrollo (ANID) of Chile.
\end{acknowledgments}


\nocite{*}

\bibliography{v1_paper_ref}

\end{document}